\documentclass[12pt,preprint]{emulateapj}
\usepackage{graphicx}
\usepackage{subfigure} 
\usepackage{apjfonts}
\usepackage{xcolor}
\newcommand{\Ia}{SN~Ia}
\newcommand{\Ias}{SNe~Ia}
\newcommand{\Msun}{~M$_{\odot}$}
\newcommand{\el}[2]{$^{#1}$#2}     
\newcommand{\vel}{~km~s$^{-1}$}   
\shorttitle{Transient Absorption Features in Late-Time Ia's} 
\shortauthors{Black, Fesen, \& Parrent}

\slugcomment{Accepted by MNRAS} 

\begin{document}

\title{Narrow Transient Absorptions in Late-Time Optical Spectra of Type Ia Supernovae:\\ Evidence for Large Clumps of Iron-Rich Ejecta?}

\author{Christine S.\ Black\altaffilmark{1}, Robert A.\ Fesen\altaffilmark{1} \& Jerod T.\ Parrent\altaffilmark{1}} 
\affil{\altaffilmark{1}6127 Wilder Lab, Department of Physics \& Astronomy, Dartmouth College, Hanover, NH 03755 }

\begin{abstract}

An examination of late-time, optical spectra of type Ia supernovae revealed surprisingly narrow absorption features which only become visible a few months after maximum light.  These features, most clearly seen in the late-time spectra of the bright, recent type Ia supernovae ASASSN-14lp and SN~2017bzc, appear as narrow absorptions at $\sim$4840 \AA, $\sim$5000~\AA, and as a sharp inflection at $\sim$4760~\AA\ on the red side of the prominent late-time  4700~\AA\ feature.  A survey of on-line archival data revealed similar features present in the spectra of ten other normal and 91T-like \Ias, including SN~2011fe. Unlike blue spectral features which exhibit progressive red-ward shifts, these narrow absorptions remain at the same wavelength from epoch to epoch for an individual SN, but can appear at slightly different wavelengths for each object. These features are also transient, appearing and then fading in one to three months.  After ruling out instrumental, data reduction, and atmospheric affects, we discuss possible explanations including progenitor mass-loss material, interaction with material from previous novae events, and absorption by large discrete clumps of high-velocity Fe-rich ejecta.
\end{abstract}


%%%%%%%%%%
% Section: Intro
%%%%%%%%%%
\section{Introduction}

The current consensus is that a thermonuclear explosion of a carbon-oxygen white dwarf in a binary system is at the heart of all type Ia supernovae (\Ias) events \citep{HoyFow60,Hil00}.  However, the nature of the progenitor system of a \Ia\ has not yet been firmly established as a variety of progenitor models can reproduce the observed explosions characteristics.   

Single-degenerate models require a CO white dwarf accreting mass from a nondegenerate companion star.  Once the white dwarf reaches the Chandrasekhar mass (M$_{\rm{Ch}}$) of 1.4\Msun, it explodes as a \Ia\ producing intermediate mass elements (IMEs) and $\sim$0.6\Msun of nickel.  An alternative single-degenerate model involves a sub-M$_{\rm{Ch}}$ white dwarf that explodes via He shell ignition which triggers the CO detonation and a subluminous explosion \citep{Nom82,HoeKho96}.  Double-degenerate models resulting in a \Ia\ invoke a system of two white dwarfs that either merge violently \citep{Kus13,Ras14} or explode when the primary white dwarf accretes the secondary \citep{Kas15}.

In addition to multiple progenitor systems, there are also a few different explosion mechanisms that can lead to a \Ia.  A pure detonation model converts the entire white dwarf mass into 1.4\Msun \el{56}{Ni} and does not produce any of the observed IMEs \citep{Arn71}.  On the other hand, a slow burning ``deflagration'' wave within the white dwarf creates the IMEs, but over produces the amount of synthesized Ni \citep{NomThiYok84}.  The more successful models invoke a deflagration to detonation transition (DDT) or ``delayed-detonation", where the explosion begins with a subsonic deflagration wave that, upon reaching supersonic speeds, leads to the detonation of the white dwarf \citep{Kho91,Hil00,Gam05}. 

While the overall photometric and spectroscopic properties are consistent across the majority of \Ias\ suggesting a repeatable explosion mechanism \citep{Bra08}, it is becoming increasingly clear that there is a larger variety of \Ias\ than originally thought. As such, a single ubiquitous progenitor channel may not be able to describe the many observed subtypes \citep{How11,Mao12,Dil12,Pat13,WanWan13}.

Two to three months after maximum brightness as the ejecta become increasingly transparent, \Ias\ transition into the so-called nebular phase when the spectra become dominated by forbidden line emission \citep{Axe80,Bow97,StrMazSol06}.  Studies of the nebular phase are somewhat limited because there are few SNe with high signal-to-noise spectra out to day +100 and fewer at even later epochs.  Additionally, the time coverage of observations is not uniform across individual SNe, producing a somewhat incomplete late-time picture.

At late times, \Ia\ spectra are dominated by strong, blue emission-like features  with the strongest feature seen at $\sim$4700~\AA\ flanked by weaker features at 4300, 5000, and 5300~\AA. Current late-time Ia models fit the nebular spectra using mainly [\ion{Fe}{2}], [\ion{Fe}{3}], [\ion{Co}{3}], and [\ion{Ni}{2}] forbidden emission lines, where the 4700~\AA\ feature has most often been viewed as a blend of forbidden Fe emission lines, including [\ion{Fe}{2}] 4814~\AA\ and [\ion{Fe}{3}] 4658, 4702, 4734~\AA\ resulting in a central wavelength of 4701~\AA\ \citep{Bow97,Maz15,Chi15,BotKas17}.

It has been known since the mid-twentieth century that the prominent late-time feature at 4700~\AA\ is not constant in wavelength, but shifts steadily towards longer wavelengths with time beginning a few weeks after maximum light \citep{Min39,Ber62,Ber65,McL63}.  Recently, this shift has been interpreted as a change in radial velocity where the [\ion{Fe}{2}] and [\ion{Fe}{3}] begin heavily blueshifted by roughly $-$4000\vel\ at earlier times \citep{Mae11,Sil13} then drift past the rest velocity to about +1500\vel at very late times \citep{Pan15,Gra15,Gra17}.  However, \citet{Fra15} claim that a change of ionization of Fe$^{++}$ to Fe$^0$ can explain the observed late-time red shifts.

\citet{Bla16} showed that the 4700~\AA\ blend is not the only feature that drifts to the red and that two adjacent features at 4850 and 5000~\AA\ move at roughly the same rate.  This suggests that a common cause is responsible for the observed redshifts of these late-time, blue spectral features.  They argued that \Ias\ as late as +300 days have not fully transitioned from the photospheric to nebular phase and that these features are the result of permitted absorptions in a blue continuum creating the appearance of `pseudo-emission' features.  

\citet{Bla16} proposed that strong permitted \ion{Fe}{2} absorptions in a receding photosphere could explain how the apparent emission features at 4700, 4850, and 5000~\AA\ seemingly shift passed their supposed zero-velocity wavelength.  This explanation is supported by spectral models that use mostly permitted Fe absorptions in combination with a forbidden line emission (\citealt{Fri14,Blo15}) or only permitted lines (\citealt{Bra08,Bra09}) to fit the data.  Such models not only fit the major and minor features in the spectra, but are adept at fitting the features as they shift.

Examination of recent, late-time spectra of \Ias, revealed a few weak, blue absorption features that do not exhibit a redward shift.  These features are relatively narrow but broader than typical interstellar medium (ISM) absorption.  A subsequent search through the published and on-line data archives revealed nearly a dozen \Ias\ with spectra taken between $+$50 and $+$200 days that show evidence of these same puzzling features, indicating that these absorption features are not especially rare.

Here we present an examination of the spectra of 12 \Ias\ in order to investigate the nature of these narrow features.  Our data set and observations are described in \S\ref{sec:14lpData}, analysis of the narrow features is described in \S\ref{sec:14lpResults}, and our discussion of these results and possible explanations are given in \S\ref{sec:14lpDisc}. In \S\ref{sec:14lpConc} we summarize our findings.

%%%%%%%%%%
% Section: Data
%%%%%%%%%%
\section{Data Set} \label{sec:14lpData}

A sample of high signal-to-noise, late-time Ia spectra was compiled using the following SN databases: the Open Supernova Catalog (OSC; \citealt{OSC}), the Weizmann Interactive Supernova Data Repository (WISeREP; \citealt{Yar12}) and the UC Berkeley Supernova Database (SNDB) which is a part of the Berkeley SN Ia Program (BSNIP; \citealt{Sil12}).  SNe were selected based on the availability of good S/N and late-time spectra at epochs between +50 and +200 days post-max.  Our archival sample contains 40 optical, late-time spectra of 10 \Ias\ and have been corrected for redshift.  In addition, observations of the bright \Ias\ ASASSN-14lp and SN~2017bzc described below, are included in this sample.

\subsection{ASASSN-14lp}\label{sec:14lpobs}

\Ia\ ASASSN-14lp in NGC 4666 (z = 0.00510) was the second brightest SN of
2014, reaching a peak magnitude of V~=~11.94 mag on 24 December 2014 with a
$\Delta$m$_{15}$ of 0.79 \citep{Sha15}.  Moderate resolution spectra of 2~\AA\ (R $\sim$
2000$-$3000) with a 0.7~\AA\ pix$^{-1}$ dispersion of ASSASN-14lp were taken using a 1.2$''$ slit, with exposure times of 2400 s and 4000 s in March and April 2015  at days $+$87, $+$114, respectively, with the MDM 2.4m telescope using the Ohio State Multi-Object Spectrograph (OSMOS).

A follow-up 900 s spectrum was obtained in May 2015 (day~$+$149) with the South African Large Telescope (SALT) using the Robert Stobie Spectrograph (RSS) with a 1.5$''$ slit and 1300 lines per mm grating for a resolution of 4~\AA\ and a dispersion of 0.8~\AA\ pix$^{-1}$.  

\subsection{SN 2017bzc}\label{sec:17bzcobs}

\Ia\ SN 2017bzc was discovered pre-maximum brightness on 7 March 2017 located at $\alpha$(2000) = 23$^{\rm h}$ 16$^{\rm m}$ 14.69$^{\rm s}$, $\delta$(2000) = $-$42$^\circ$ 34$^{\prime}$ 10${\farcs}$9 in NGC 7552 by the Backyard Observatory Supernova Search (BOSS) \citep{Par17,MorSha17}.  It is situated to the northeast of the galaxy center, outside of the host galaxy in a relatively uncontaminated region.  Optical spectra with 5~\AA\ resolution and a dispersion of 1.0~\AA\ pix$^{-1}$ were taken with the Robert Stobie Spectrograph (RSS) at SALT with grating of 900 lines per mm and 1.5$''$ longslit.

We adopt 15 March 2017 as the date of maximum brightness with a V$_{max}$ $\simeq$ 12.2 mag.  Observations were made at day $+$103, $+$107, $+$133, $+$165, and $+$239 spanning the wavelength range 3200 $-$ 6400~\AA\  and at days $+$102, $+$132, and $+$164 for the 6000 $-$ 9200~\AA\ part of the spectrum with exposure times ranging from 600 to 1200 s for both red and blue regions. Two tilts were done at each epoch (except day $+$239) in order to cover the chip gaps.

Data reduction for both SNe
was done using IRAF\footnote{IRAF is distributed by the National Optical
Astronomy Observatories, which are operated by the Association of Universities
for Research in Astronomy, Inc., under cooperative agreement with the National
Science Foundation.} and consisted of bias and background subtraction,
wavelength calibration, aperture extraction, and host galaxy redshift
correction.

%%%%%%%%%%
% Section: Results
%%%%%%%%%%
\section{Results}\label{sec:14lpResults}

%%%%%%%%%%
% Figure: SN 2017bzc
%%%%%%%%%%
\begin{figure*}
        \centering
        \includegraphics[width=0.78\textwidth]{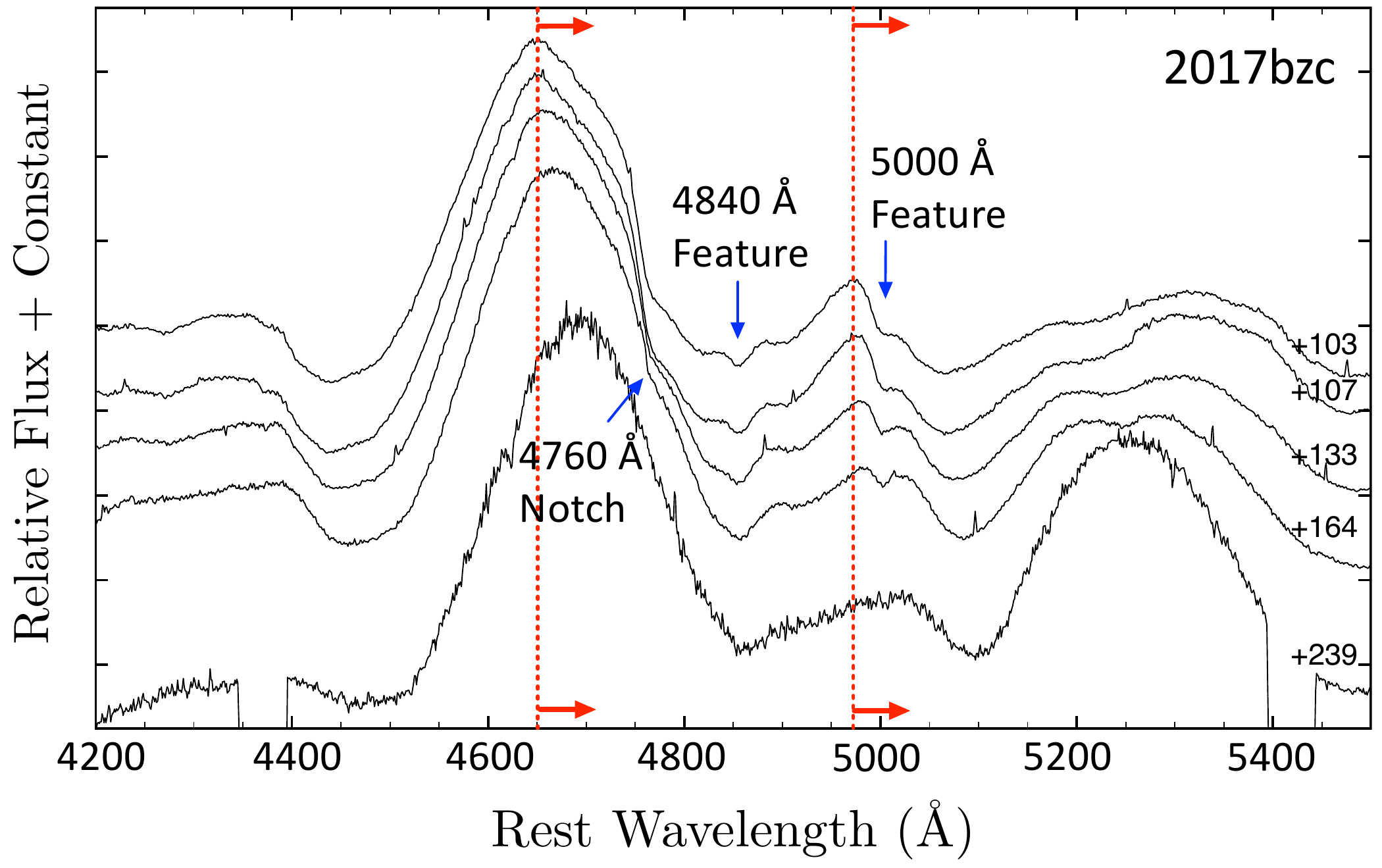}
        \caption[SALT Spectra of SN~2017bzc with the Features Marked by Arrows]{The SALT spectra of SN~2017bzc with the three narrow features marked by blue arrows.  The red dotted lines mark the central wavelengths of the pseudo-emission features at 4700 and 5000~\AA\ at day +103 and the red arrows indicates that these features shift to longer wavelengths.}
        \label{fig:17bzc}
\end{figure*}
%%%%%%%%%%

%%%%%%%%%%
% Subsection: Narrow Absorption
%%%%%%%%%%
\subsection{Narrow, Late-Time Absorption Features}

During our analysis of late-time \Ias\ spectra \citep{Bla16}, examination of the near and bright \Ia\ ASASSN-14lp revealed some unexpectedly narrow absorption features.  These include a small narrow absorption at $\sim$5000~\AA, a weak absorption at $\sim$4840~\AA, and a sharp inflection or `notch' at $\sim$4760~\AA, on the red side of the prominent late-time 4700~\AA\ feature. Subsequent examination of another recent bright, late-time spectra of SN~2017bzc showed similar features.  

Figure \ref{fig:17bzc} presents high S/N spectra of SN~2017bzc covering the region 4200 - 5500~\AA\ at day +103, +107, +133, +164, and +239.  The red dotted lines mark the approximate central wavelengths of the 4700 and 5000~\AA\ emission features at day +103 with red arrows indicating their progressive redward shift with time. 

As can be seen in Figure \ref{fig:17bzc}, these three weak absorptions do not shift with time.  This is unlike the strong 4700~\AA\ and several blue spectral features which show progressive red shifts of 40~\AA\ or more at late times \citep{Bla16}.  

The stationary nature of these absorption features is especially apparent in the 5000~\AA\ region where the weak but broad emission feature shifts redward ``underneath'' the narrow absorption.  Consequently, the absorption appears on the red side of the 5000~\AA\ emission at day +103, but by days +133 and +164 the absorption is nearly centered on the emission.  Examination of spectra covering 6000 - 9200~\AA\ did not reveal any similar narrow and stationary features.

%%%%%%%%%%
% Figure: 14lp, 17bzc, 94D, 11fe
%%%%%%%%%%
\begin{figure*}
        \centering
        \includegraphics[width=0.8\textwidth] {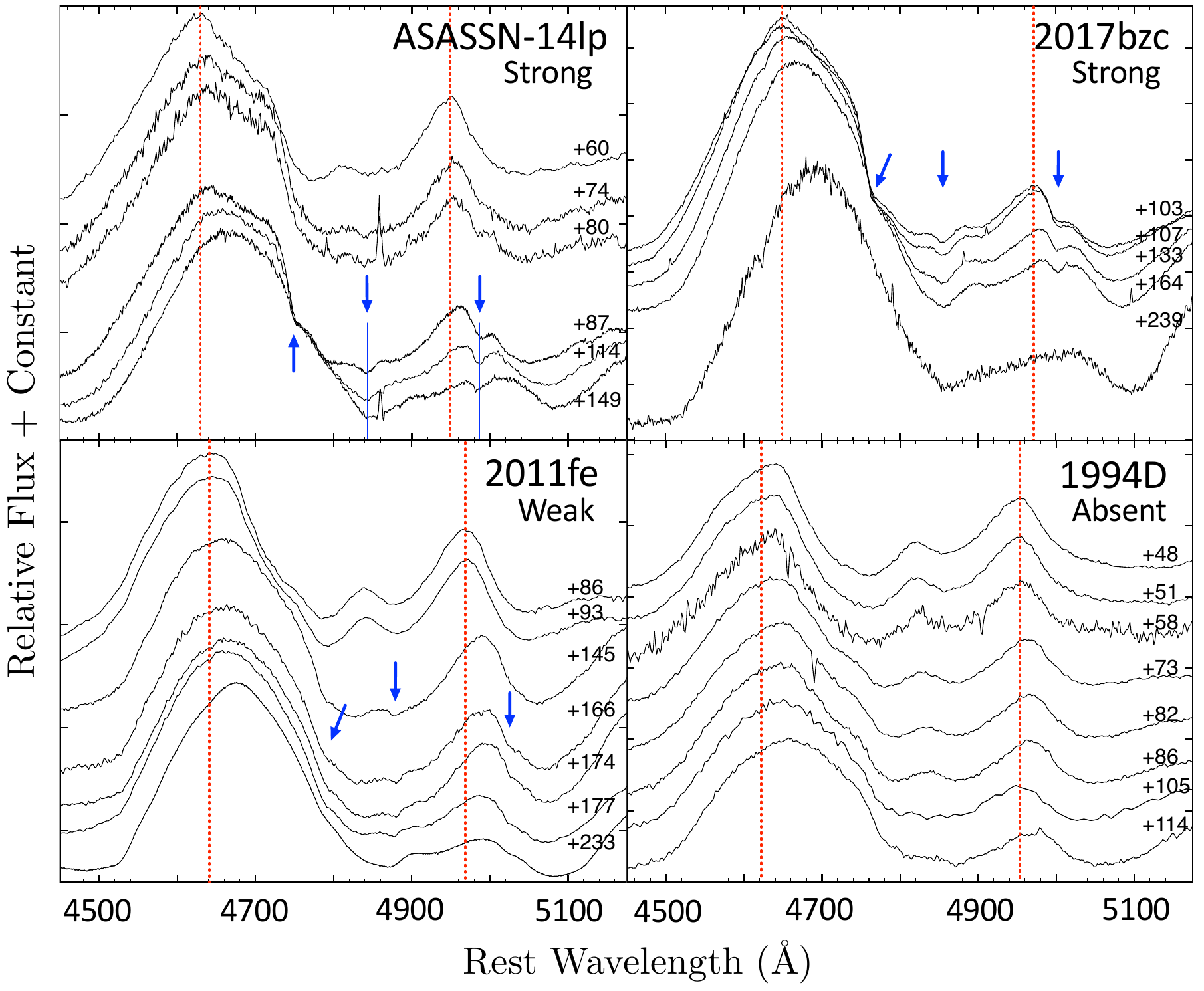}
        \caption{Top: Spectra of ASASSN-14lp and SN 2017bzc where the spectra are stacked at the 4760~\AA\ notch.  The spectra of ASASSN-14lp at days +60, +74, and +80 are from \citet{Sha15}.  Bottom: Spectra of the normal SN~1994D for comparison and SN~2011fe, which also exhibits these features.  Blue solid lines and arrows mark the wavelengths of the minor features while red dotted lines are the same as those in Figure \ref{fig:17bzc}.  Note how the minor features do not appear to change in wavelength and the absorption at 4760~\AA\ appears to be cutting in to the 4700~\AA\ `emission' feature.}
        \label{fig:SNcompare}
\end{figure*}
%%%%%%%%%%

%%%%%%%%%%
% Figure: SN Sample
%%%%%%%%%%
\begin{figure*}[!t]
        \centering
        \includegraphics[width=0.9\linewidth]{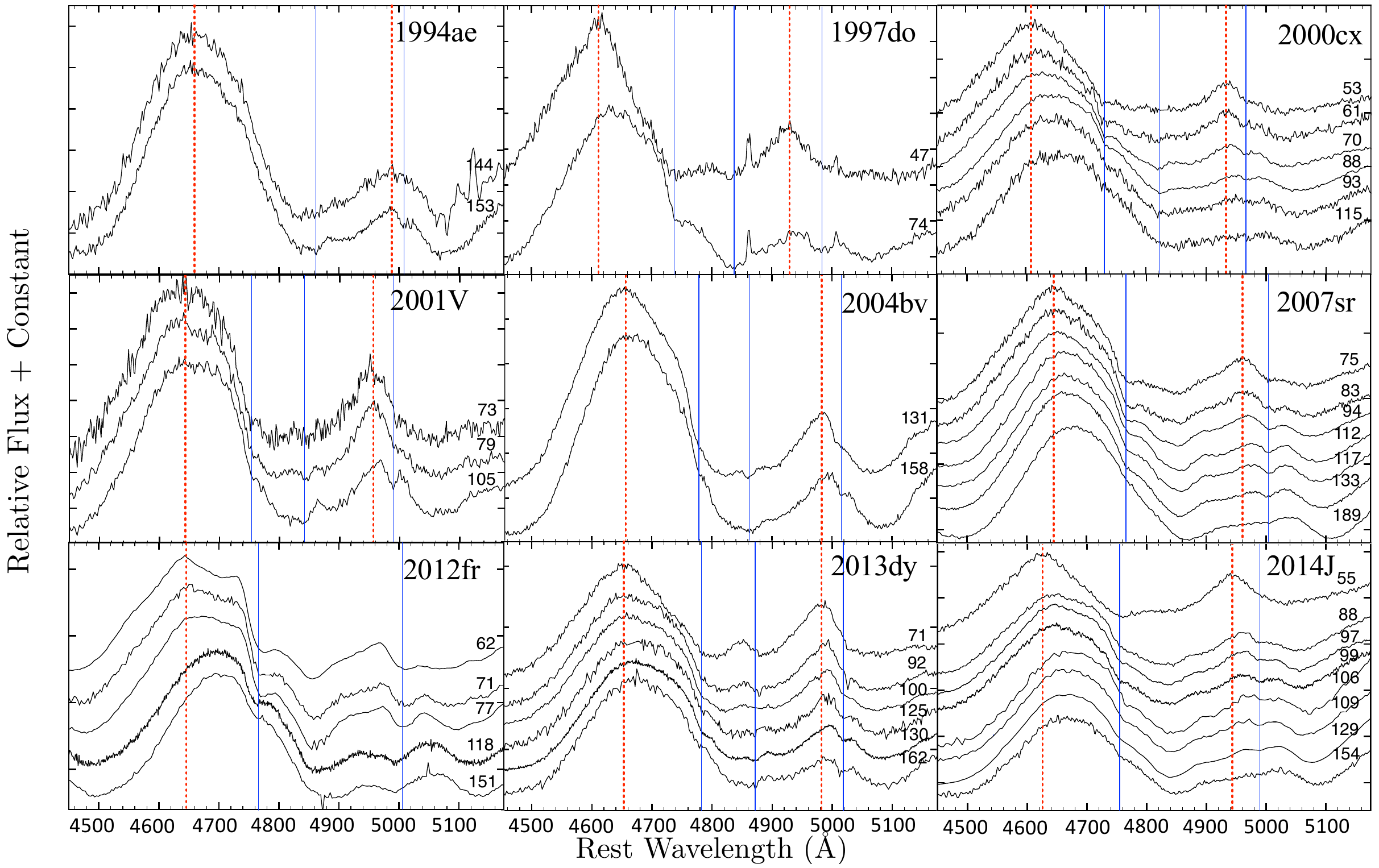}
        \caption{Nine \Ias\ from the online archives that also show these narrow absorptions.  The red dotted and blue solid lines are the same as those in Figure \ref{fig:SNcompare}.  All epochs are listed as days post-maximum brightness.}
        \label{fig:allSN}
\end{figure*}
%%%%%%%%%%

%%%%%%%%%%
% Table: Delta m15 for SN  in Fig.
%%%%%%%%%%
\begin{deluxetable*}{lcccl}
  \tablecolumns{5}
  \tablecaption{\Ias\ Showing Late-Time Narrow Absorption Features}
  \tablehead{
  \colhead{SN} & \colhead{m$_{max}$(B)} & \colhead{$\Delta$m$_{15}$ (B)} & \colhead{Subtype} & \colhead{References}}
  \startdata
  	1994ae & 13.21 & 0.86 & normal & \citet{Rie99} \\
	1997do & 14.56 & 0.99 & normal &  \citet{Jha0697do} \\
	2000cx & 13.43 & 0.93 & 91T-like/pec & \citet{Li01,Blo12} \\
	2001V & 14.60	& 0.9-0.99	& 91T &  \citet{Vin03} \\
	2004bv & 13.89 & 0.88 & normal &  \citet{Gan10} \\
	 & & & & \citet{Sil12} \\
	2007sr & 12.71 & 1.05-1.13 & normal &  \citet{Gan10} \\
	 & & & & \citet{Blo12} \\
	2011fe & 9.9 & 1.05 & normal & \citet{Mun13,Maz14}\\
	2012fr & 12.0 & 0.80-0.85 & 91T-like &  \citet{Chi13,Zha14}  \\
	 & & & & \citet{Sil15} \\
	2013dy & 13.28 & 0.92 & normal &  \citet{Zhe13,Pan15}\\
	20014J & 10.61$^a$ & 1.01-1.12 & normal & \citet{Fol14,Gal16}  \\
	ASASSN-14lp & 11.94$^a$ & 0.79 & normal &  \citet{Sha15}\\
	2017bzc & 12.2$^a$ & unknown & unknown &  \citet{Par17}
   \enddata
   \label{tab:Sample}
   \tablenotetext{a}{The reported magnitudes for both SN~2014J and ASASSN-14lp are in V; SN~2017bzc is filterless.}
\end{deluxetable*}
%%%%%%%%%%

%%%%%%%%%%
% Figure: SN Stack
%%%%%%%%%%
\begin{figure}[!t]
        \centering
          \includegraphics[width=\columnwidth]{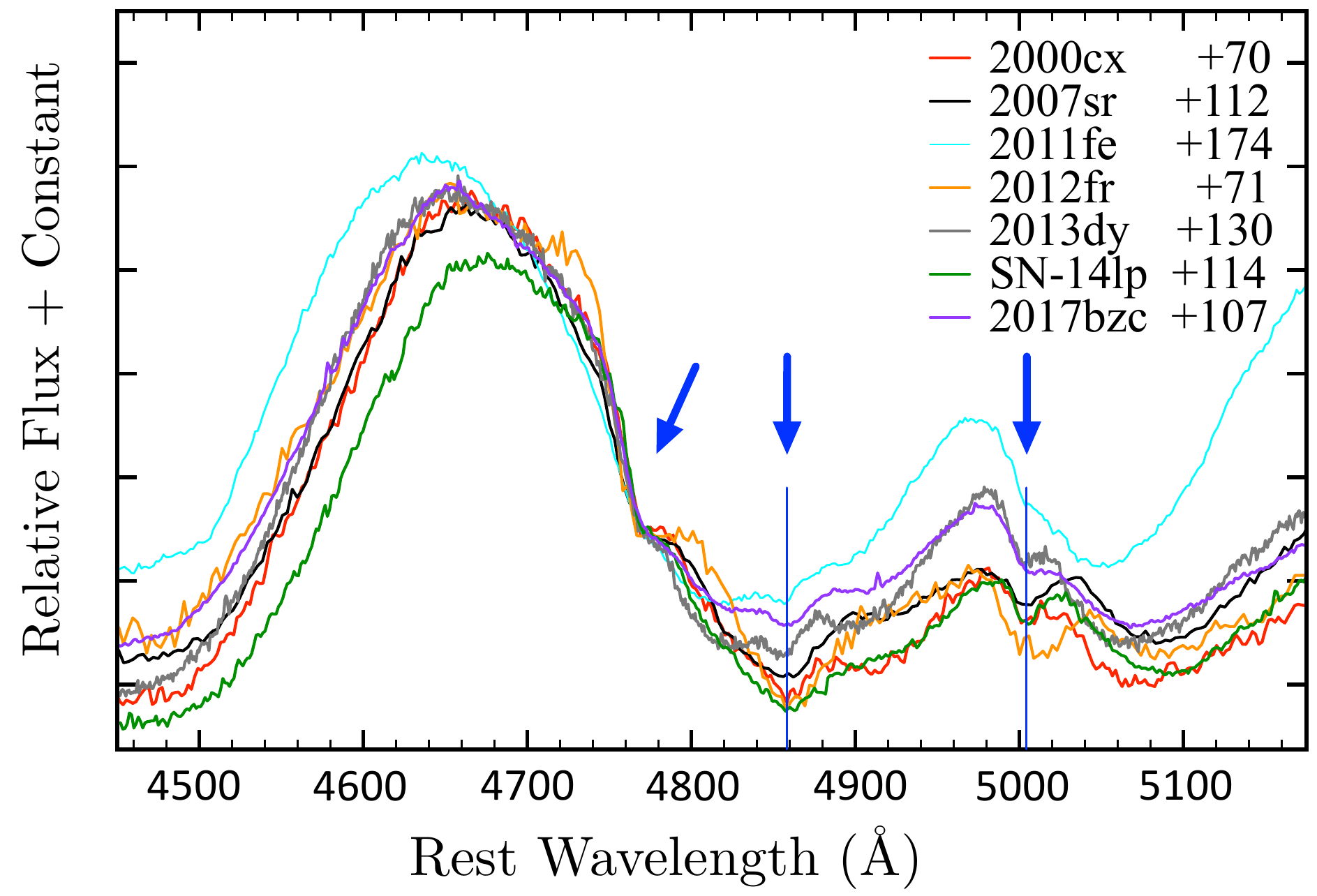}
        \caption[Seven \Ias\ Stacked at their 4760~\AA\ Notch]{Seven \Ias\ stacked such that their 4760~\AA\ notch features are aligned with the notch seen in SN~2017bzc.}
        \label{fig:SNstack}
\end{figure}
%%%%%%%%%%

%%%%%%%%%%
% Figure: Wavelength Plots
%%%%%%%%%%
\begin{figure*}
        \centering
          \subfigure[]{ \label{fig:4700knick}
		\includegraphics[width=\columnwidth]{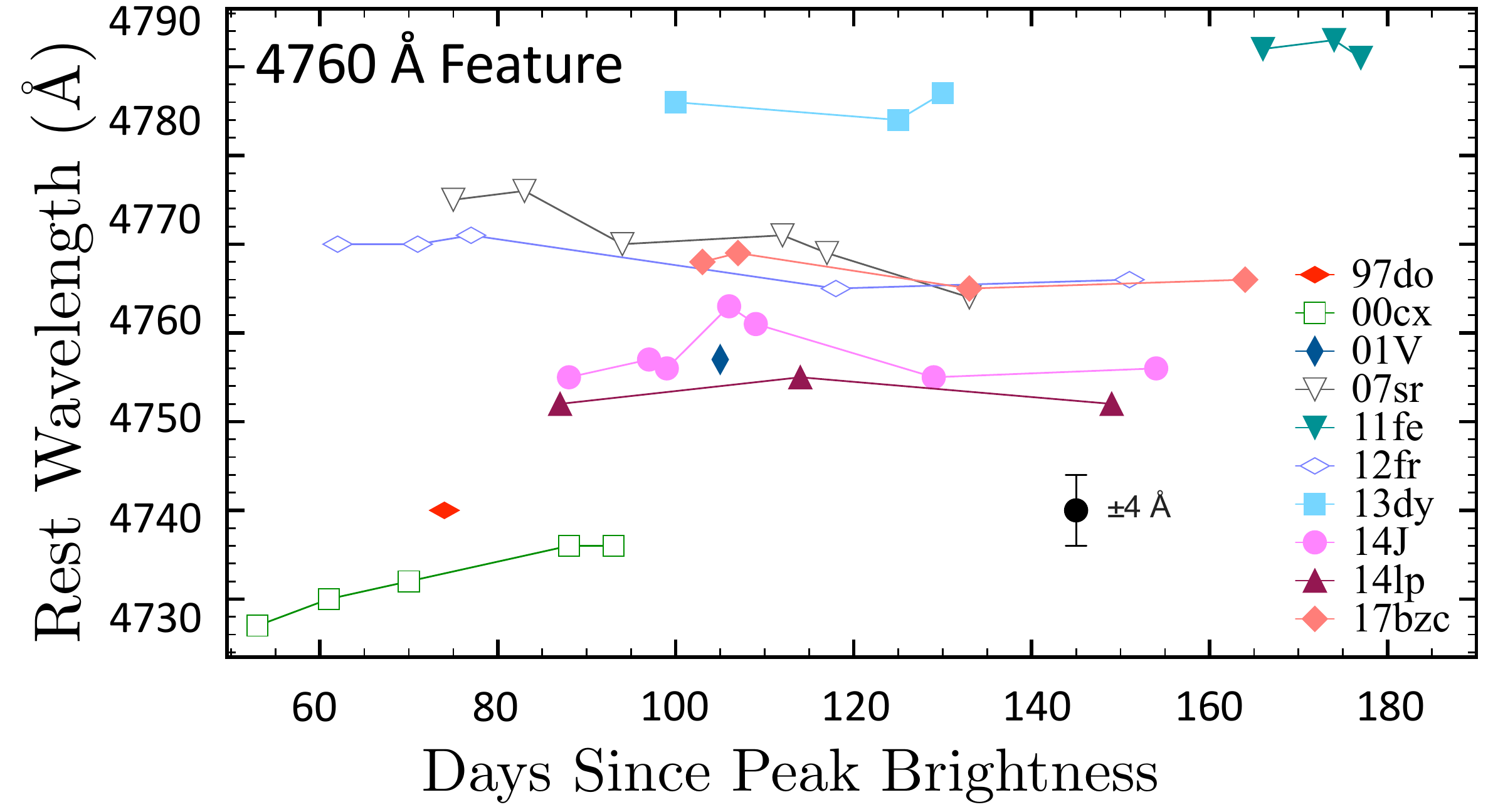}}
          \subfigure[]{ \label{fig:5000small}
		\includegraphics[width=\columnwidth]{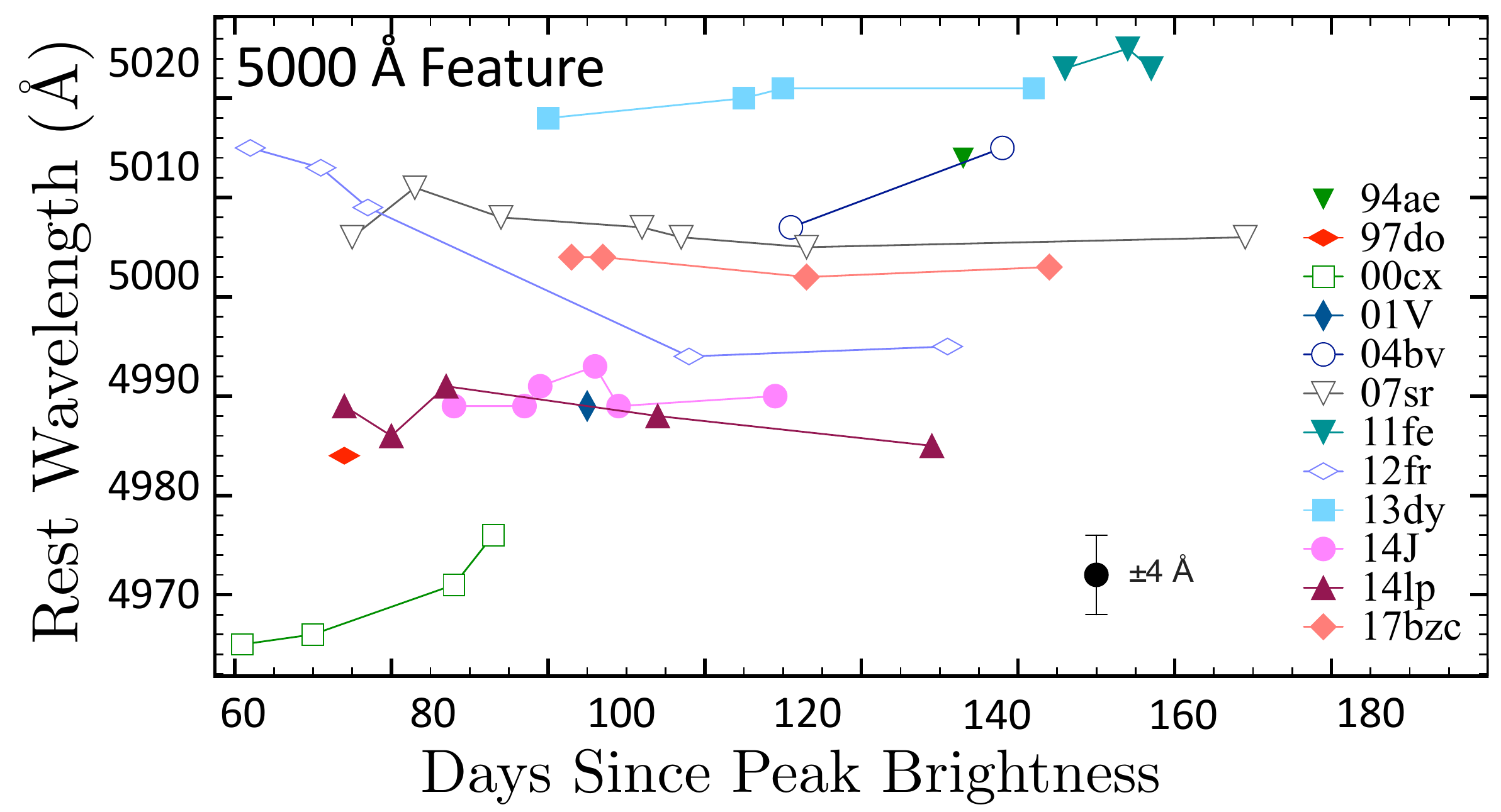}}
          \subfigure[]{ \label{fig:Sep5047}
		\includegraphics[width=\columnwidth]{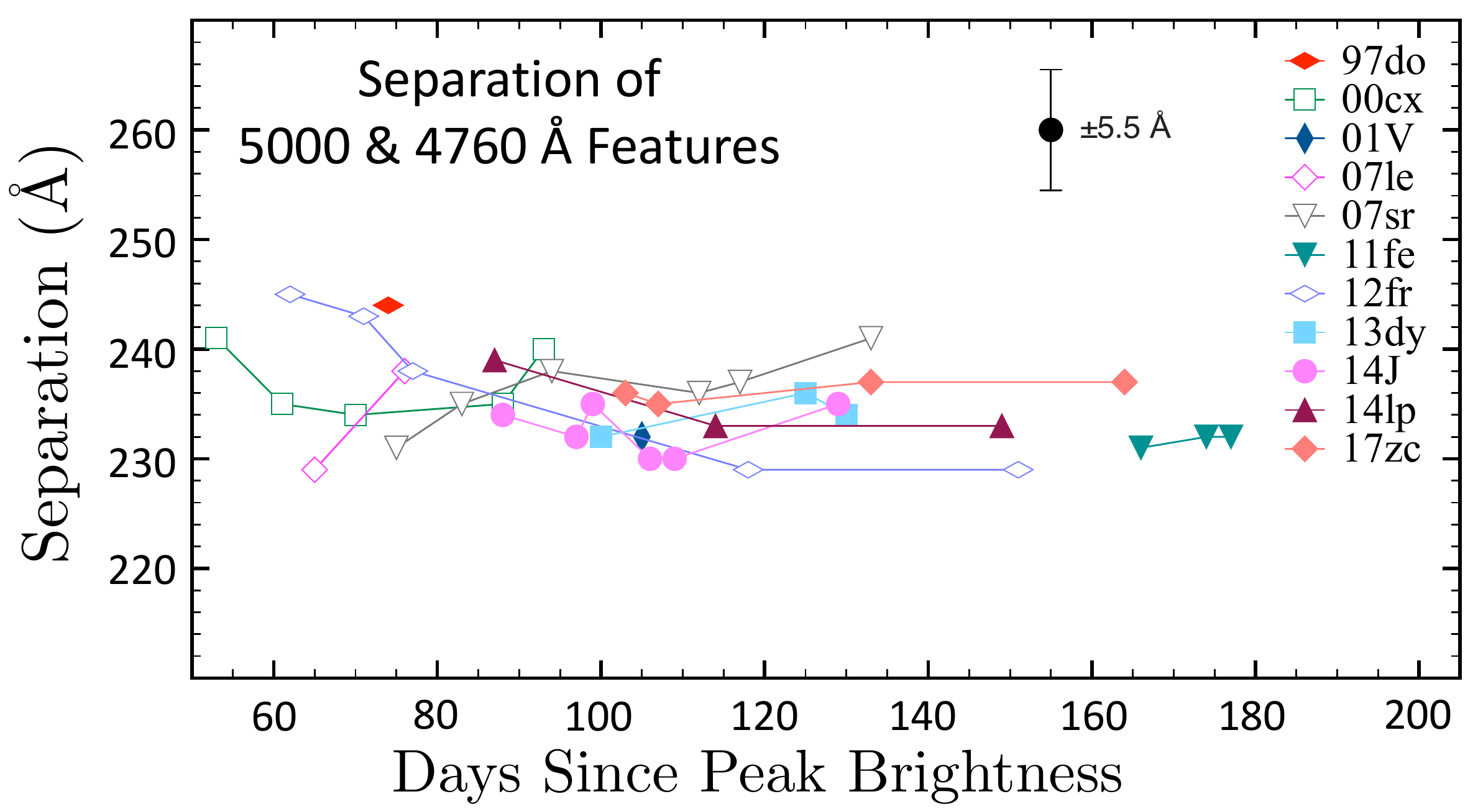}}
          \subfigure[]{ \label{fig:Sep5048}
		\includegraphics[width=\columnwidth]{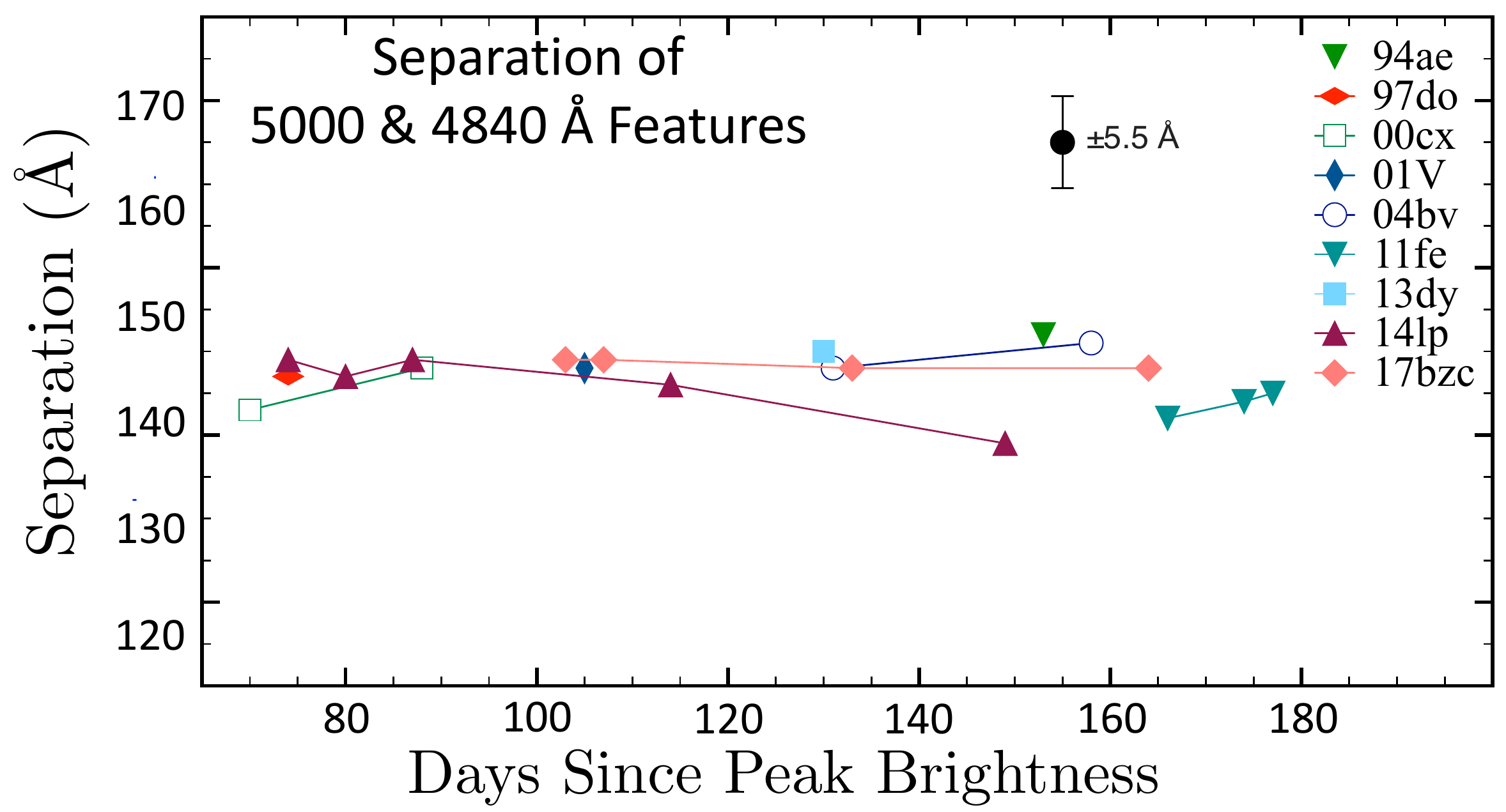}}
        \caption{The approximate wavelength of the narrow absorption features seen at (a) $\sim$4760~\AA, (b) $\sim$4840~\AA\ and the separation of the (c) 5000 and 4760~\AA\ features and (d) 5000 and 4840~\AA\ features across all \Ias\ in our sample.}
        \label{fig:waveplot}
\end{figure*}
%%%%%%%%%%

To illustrate that these features maintain constant wavelengths, Figure \ref{fig:SNcompare} shows the spectra of ASASSN-14lp and SN~2017bzc stacked with flux normalized to the 4760~\AA\ notch. The blue arrows indicate the three narrow features, while the thin blue solid lines mark the approximate centers of the 4840 and 5000~\AA\ absorptions, with the red dotted lines the same as those in Figure \ref{fig:17bzc}.  As can be seen, all three features - 4760, 4840, and 5000 \AA\ - show no change in wavelength.

Interestingly, late-time spectra of the bright \Ia\ SN~2011fe (bottom right panel) shows indications that these absorptions may also be present.  However, unlike ASASSN-14lp and SN~2017bzc, the features in SN~2011fe are much weaker and appear at later epochs (See Fig. \ref{fig:SNcompare}). For comparison, in bottom left panel we show the spectra of the normal SN~1994D, which does not show these narrow absorptions.

In the spectra of ASASSN-14lp and SN~2017bzc, the wavelength of the 4760~\AA\ notch is strikingly consistent, even at epochs 60 days apart.  While the strong 4700~\AA\ emission feature shifts steadily towards the red, the notch on the red side of the 4700~\AA\ profile remains remarkably constant in wavelength.  This is especially apparently in the spectra of ASASSN-14lp.

A subsequent search of on-line SNe spectral catalogues found nine additional \Ias\ which exhibit evidence for these same features (see Table \ref{tab:Sample}). Figure \ref{fig:allSN} shows the spectra of the other nine SNe that present narrow absorptions, where the dotted and solid lines are the same as those seen in Figure \ref{fig:SNcompare}.  We note that a few additional \Ias, including 1995al, 1999ac, and 2007gi, may also exhibit similar absorptions, but are excluded due to weak detections.

In Figure \ref{fig:SNstack} we show the spectra of seven \Ias\ (2000cx, 2007sr, 2011fe, 2012fr, 2013dy, ASASSN-14lp, and 2017bzc) stacked such that the wavelengths of the 4760~\AA\ notches are aligned with that of SN~2017bzc, thus ignoring wavelength differences exhibited by the individual SNe. Plotted in this way, the three narrow features seen in these six different \Ias\ all exhibit virtually identical separations from one another, emphasized by the blue line which denotes the center wavelengths of the features in SN~2017bzc.  This indicates that the relationship between these absorptions is repeatable, including the case of the unusually broad absorption features in SN~2012fr.

The width of the 5000~\AA\ feature varies amongst our SNe sample (see Fig. \ref{fig:SNstack}).  We find a range in FWHM for the 5000~\AA\ feature of roughly 10 to 35~\AA\ and does not appear to be correlated to the wavelength of the feature.  While the absorption at $\sim$4840 \AA\ is distinct in some \Ias, like ASASSN-14lp and SN~2017bzc, it is not seen in the entire sample and only definitively visible in nine SNe in our sample. 

A noticeable difference between \Ias\ that show these features and those that do not, such as the prototypical \Ia\ 1994D (Fig. \ref{fig:SNcompare}), is the epoch at which the broad, but weak emission feature at $\sim$4850~\AA\ disappears. It was shown in \citet{Bla16} that this feature fades in nearly all \Ias, with SN~2011fe being an exception (see Fig. 5 of \citealt{Bla16}).  It can be seen in Figures \ref{fig:SNcompare} and \ref{fig:allSN} that the emission feature at 4850~\AA\ disappears from the spectra of the ``14lp-like'' SNe at generally earlier epochs ($\sim$ +60-70 days) than the 94D-like SNe ($\sim$ +100 day).  The feature lingers much longer in SN~2011fe, making it a clear outlier among the SNe that exhibit these features.

Since high S/N, high cadence spectral data of \Ias\ past day +100 are still relatively uncommon, most of the time coverage in our sample do not show the full evolution of these features.  Some SNe, like 2001V, only have one or two spectra that show evidence of the weak absorptions.  Other \Ia, like 2012fr, have a number of spectra and are more extreme examples with unusually strong and broad 4850 \& 5000~\AA\ features.  

%%%%%%%%%%
% Subsection: Transient Nature
%%%%%%%%%%
\subsection{Feature Wavelengths and Separations with Time}

Examination of the available spectra reveals that, from object to object, all three features maintain nearly constant wavelengths to within measurement error ($\pm$4~\AA). The exception is SN~2012fr.  The 5000~\AA\ feature in SN~2012fr exhibits an apparent blueshift of 15-20~\AA\ between days +77 and +118.

Figures \ref{fig:4700knick} and \ref{fig:5000small} plots the wavelengths of the 4760 and 5000~\AA\ features for each \Ia\ versus time.  Additionally, the absorptions exhibit roughly similar central wavelengths across our sample, but with a range of up to $\pm$30~\AA.  This consistency throughout our sample suggests a common cause connecting these three features.

Figures \ref{fig:Sep5047} and \ref{fig:Sep5048} depict the separations between the 5000 and 4760~\AA\ features and the 5000 and 4840~\AA\ features near day +100, respectively.   These narrow absorptions are found to exhibit the same separations within a few Angstrom of each other for nearly every \Ia\ in the sample, where SN~1997do is the single outlier.  Even the features in the extreme case of SN~2012fr fall within these separations as well as SN~2011fe despite its features found higher wavelengths.

%%%%%%%%%%
% Subsection: Transient Nature
%%%%%%%%%%
\subsection{The Transient Nature of these Features}

The spectra of \Ia\ in the nebular phase evolve relatively slowly.  In contrast, however, these narrow absorptions appear and then disappear in a matter of just 1 $-$ 3 months.  For a majority of SNe, the narrow absorptions are present in the spectra roughly simultaneously, including the 4840~\AA\ feature.  A few \Ia\ do not show the 4760~\AA\ notch, though this may be the result of low S/N spectra since the initial appearance of the feature can be subtle.

%%%%%%%%%%
% Figure: Time Hist
%%%%%%%%%%
\begin{figure}
        \centering
		\includegraphics[width=0.9\columnwidth]{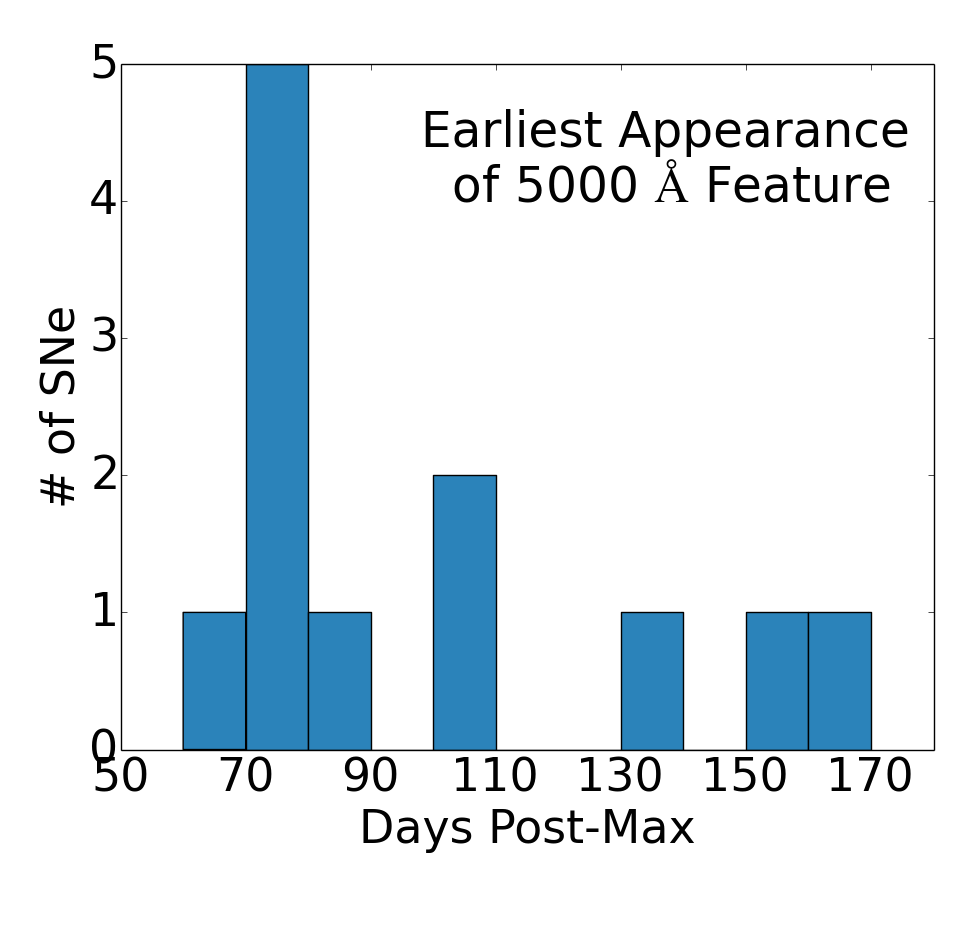}
        \caption[First Epoch where Features are Observable]{Day where the 5000~\AA\ feature is first visible in the available data.}
        \label{fig:Time_hist}
\end{figure}
%%%%%%%%%%

In most of our 12 SNe, the features are initially present between +60 and +90 days, though they can appear later than day +130, as seen in SNe~1994ae, 2004bv, and 2011fe.  Three SNe in the sample have spectra that show the absorptions' entire evolution, SNe~2000cx, 2007sr, and 2014J, where the features are seen to last for roughly 30 to 60+ days.  Spectra of SN~2014J at days +106, +109, and +129 are from \citet{Zha18}.  In the case of SN~2012fr, despite lacking full time coverage, the features remain in the spectra for nearly 100 days.  The length of time the features are visible varies from SNe to SNe and these absorptions can be seen for at least 1 $-$ 3 months before disappearing from the spectra.
  
Figure \ref{fig:Time_hist} shows the earliest appearance of the 5000~\AA\ in the available spectra, but due to gaps in the late-time observations these values reflect the earliest day that the features can be seen in the available data, not necessarily the epoch they first appear.

%%%%%%%%%%
% Subsection: Sample Distribution
%%%%%%%%%%
\subsection{Distribution Across \Ias\ Subtypes}

\begin{figure}
        \centering
          \subfigure[]{ \label{fig:5000-Type}
		\includegraphics[width=0.9\columnwidth]{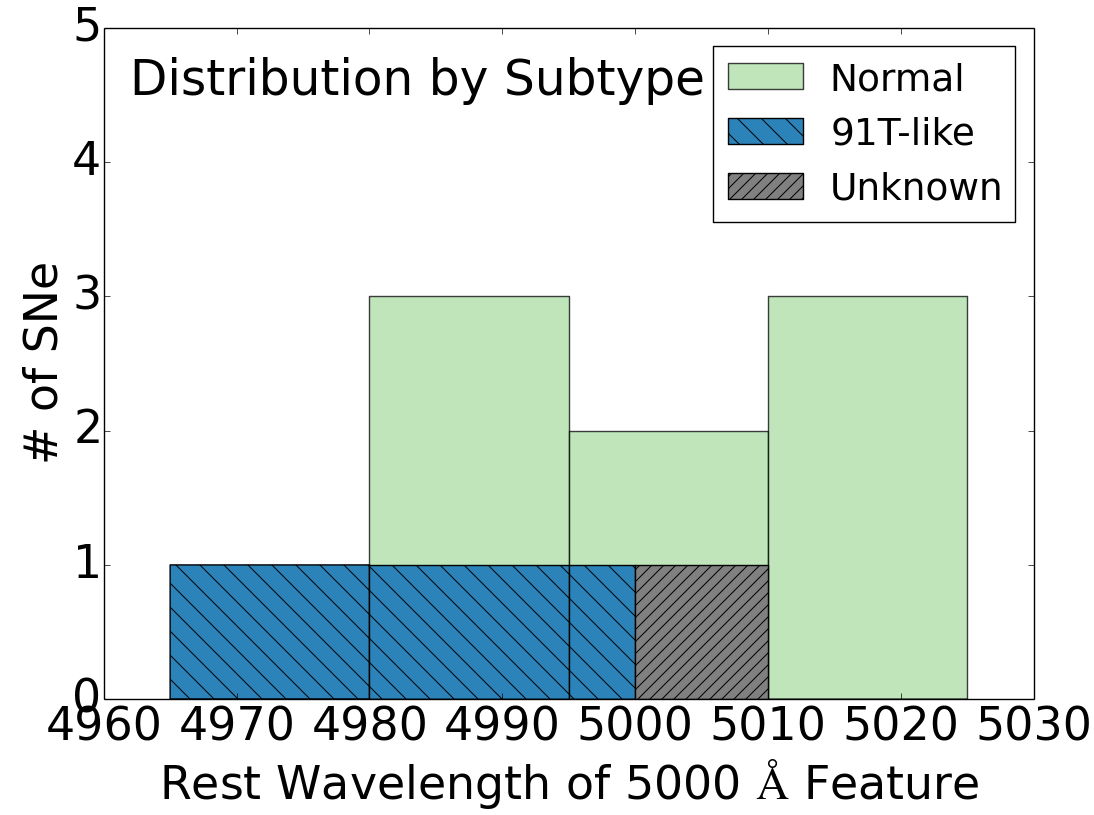}}
          \subfigure[]{ \label{fig:5000-m15}
		\includegraphics[width=0.9\columnwidth]{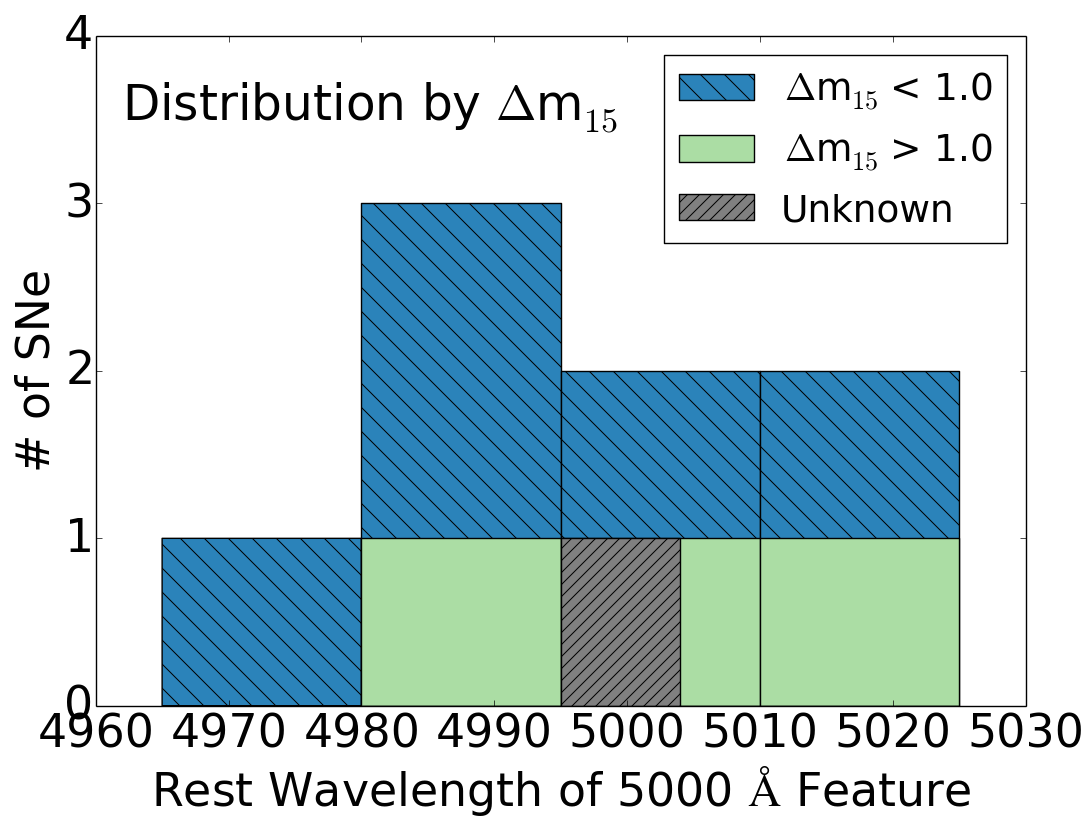}}
          \subfigure[]{ \label{fig:redshift}
		\includegraphics[width=0.9\columnwidth]{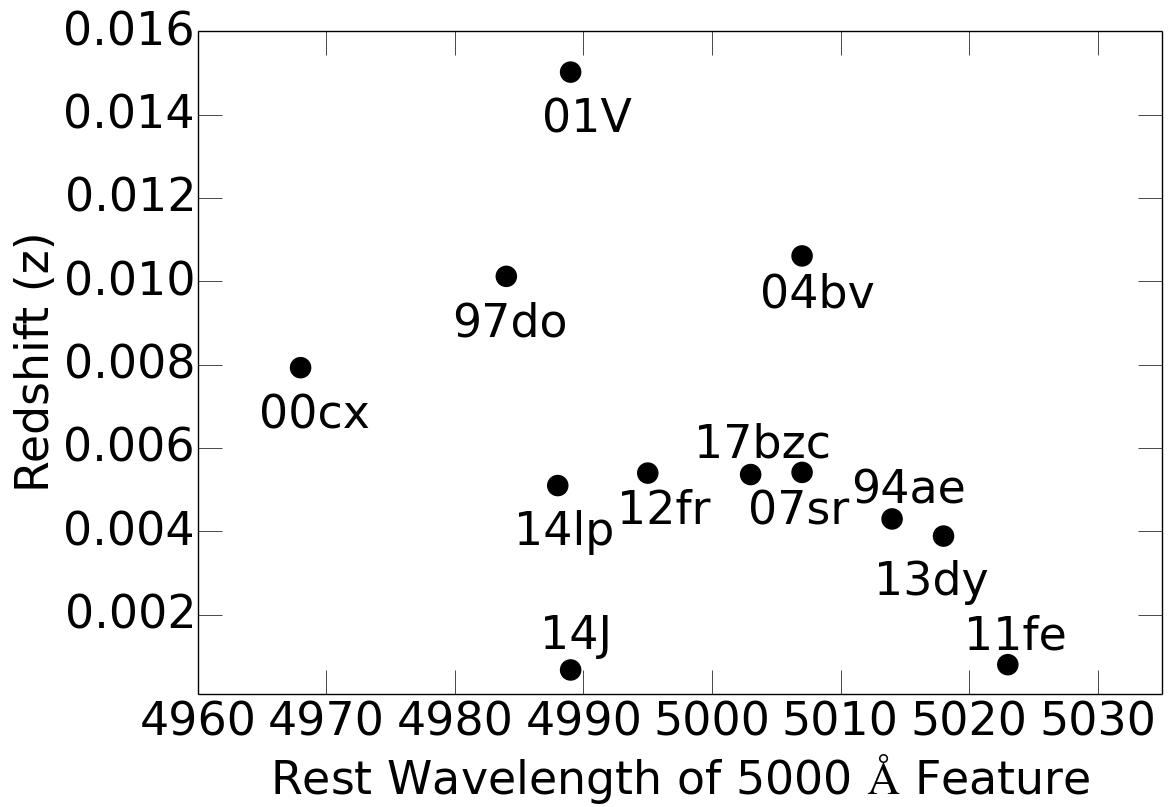}}
        \caption[Distribution of SNe in Our Sample Based on $\Delta$m$_{15}$, Type, and Redshift]{The distribution of \Ias\ in our sample (a) $\Delta$m$_{15}$, (b) SN type, and (c) redshift based on the approximate central wavelength of the narrow absorption feature at $\sim$5000~\AA.}
        \label{fig:Distrib}
\end{figure}

Our sample includes both normal and 91T-like SNe.  Table \ref{tab:Sample} lists the SNe in our sample with their peak B magnitude, $\Delta$m$_{15}$, and their \Ia\ classification.  To date, no near max spectra or photometry have been published for SN~2017bzc, so in this work its $\Delta$m$_{15}$ and classification are listed as `unknown' in Table \ref{tab:Sample} and Figure \ref{fig:Distrib}.

%%%%%%%%%%
% Figure: Histograms
%%%%%%%%%%
%%%%%%%%%%

These narrow absorptions are not limited to any one \Ia\ subtype as they appear in both normal and 91T-like SNe.  Figure \ref{fig:5000-Type} shows how the SN type relates to the wavelength of the 5000~\AA\ feature near day +100.  There appears to be a trend where 91T-like SNe have a bluer 5000~\AA\ feature than normal Ia's, though a larger sample is needed to be certain.

Figure \ref{fig:5000-m15} displays the distribution of \Ias\ based on the $\Delta$m$_{15}$ above and below 1.0.  It should be noted that all \Ias\ in our sample have $\Delta$m$_{15}$ values $<$ 1.13.  This suggests that these absorptions may be more common in slow declining SNe, but we caution that this may be the result of a selection effect due to their generally higher luminosities.

There does not appear to be a correlation between the wavelength of the 5000~\AA\ absorption and the redshift of the SNe (see Figure \ref{fig:redshift}).  This suggests that the variations in wavelengths of the narrow features observed here is not caused by the inherent redshift of higher z events and is unique to each \Ia.

%%%%%%%%%%
% Subsection: SN Fraction
%%%%%%%%%%
\subsection{Percentage of \Ias\ with Narrow Absorptions}

A search of the SNe archives revealed about 100 \Ias\ that have spectra of sufficient S/N between +50 and +200 days to clearly show these narrow absorptions.  In addition to our spectra of ASASSN-14lp and SN~2017bzc, we found nine out of the 100 SNe exhibit these features ($\sim$10\%), suggesting that the presence of narrow absorption in late-time \Ia\ spectra is not particularly uncommon.

It is interesting to note that since 2010 there have been 12 \Ias\ that reached peak magnitudes of +12.5 or brighter and of these four (30\%) exhibit these narrow, late-time absorptions (See Table \ref{tab:Sample}). In addition, if the spectra of a prototypical \Ia\ like SN~2011fe exhibit these transient absorptions at later times (day +150 or later), then this suggests that these features are not rare and may be more easily seen if high S/N spectra are obtained at late times.  

%%%%%%%%%%
% Section: Discussion
%%%%%%%%%%
\section{Discussion}\label{sec:14lpDisc}

Late-time optical spectra of \Ias\ follow a well established pattern of declining flux dominated by numerous broad features, strongest in the blue.  In Section \ref{sec:14lpResults}, we presented late-time spectra of nearly a dozen \Ias\ that show surprisingly narrow features, appearing between 50 and 200 days post-max. These transient absorptions are found at $\sim$4840 and 5000~\AA, as well as a notch-like feature on the red side of the 4700~\AA\ feature at roughly 4760~\AA.

Possible explanations concerning the origins of these absorptions must address the following observational properties: 1) The features appear simultaneously in late-time \Ia\ spectra and show no change in wavelength, unlike the prominent blue spectral features which exhibit considerable and progressive red shifts, 2) They do not appear at precisely the same wavelengths from object to object, yet the relative spacing between the narrow features remains virtually constant at all epochs within measurement error, 3) These absorptions are transient, becoming visible around day 60 and persist for one to three months before disappearing, and 4) They are seen in both normal and 91T-like \Ias.

Below we discussed several possible causes.

%%%%%%%%%%
% Subsection: Possible answers
%%%%%%%%%%

%%%%%%%%%%
% Subsection: What it's not
%%%%%%%%%%
\subsection{Instrumentation, Telluric, and Data Reduction Effects}

The weakness of these features plus the lack of other narrow absorptions in either the red or other blue parts of the spectrum raises the possibility that these features are instrumentation effects.  This explanation can be ruled out since these features do not appear in the spectra of standard stars.  Additionally, they would not appear only in spectra spanning a few months.  A variety of instruments were used to observe the SNe in our sample and these same instruments have also been used to collect data of \Ias\ that do not show these minor features.

We can also rule out a telluric origin since the absorption features should always be present and found at the exact same wavelengths.  Moreover, telluric absorptions are strongest in the red and NIR parts of the spectrum  where we see no such features in our sample.

Because the 4840 and 5000~\AA\ absorptions are at wavelengths close to H$\beta$ and [\ion{O}{3}] 5007~\AA, we also considered the possibility of over-subtraction of nearby H~II region emission during data reduction.  However, over-subtracted H$\beta$ or [\ion{O}{3}] emission would create a feature that is shifted only a few angstroms from its rest wavelength and not found red-ward of the H$\beta$ and [\ion{O}{3}] rest wavelengths.

%%%%%%%%%%
% Subsubsection: ISM
%%%%%%%%%%
\subsection{Interstellar Absorption}

These features could be due to interstellar medium (ISM) contamination in the spectra, which can be detected, for example, by narrow \ion{Na}{1} absorption or diffuse interstellar bands (DIBs).  \citet{Phi13} found ISM and DIB features present in 32 \Ias\ by way of narrow absorptions of \ion{Na}{1}~D 5890, 5896~\AA, \ion{K}{1} 7665, 7699~\AA, and the DIB feature at 5780~\AA.  Interestingly, SN~2007sr is also included in their sample, but the narrow absorptions at 4770 and 5006~\AA\ were not discussed.  We noted that varying DIB absorptions have also been seen in the Type Ic SN~2012ap \citep{Mil14}.

However, the wavelengths of the 4760, 4840, and 5000~\AA\ features described above do not correspond to any strong, known DIB or ISM species.  Additionally, the features we have reported are both variable in wavelength, with a range of wavelengths of up to 20~\AA, and are too broad to be likely ISM features.  For example, the 5000~\AA\ absorption in ASASSN-14lp and SN~2017bzc have a FWHM of roughly $\sim$12~\AA, which is much larger than the $<$1~\AA\ usually seen in the ISM.  Measured FWHM values of the 5000 \AA\ absorptions for our sample are listed in Table \ref{tab:Vel}.

%%%%%%%%%%
% Subsubsection: HV Material
%%%%%%%%%%
\subsection{Detached High-Velocity Ejecta Material}

We also considered the possibility that these features are the result of large blueshifted absorptions from high-velocity SN ejecta.  High-velocity material has been detected chiefly in \ion{Si}{2} and \ion{Ca}{2} absorptions near maximum, lying above the photosphere and `detached' from the SN's photospheric features.  These detached absorption features have velocities of 20,000 to 30,000\vel and have been seen in a handful of \Ias, including SN 1999ee \citep{MazBenSte05}, SN 2000cx \citep{Li01, Tho04}, SN 2001el \citep{Wan03}, and SN 2006D \citep{Tho07}.

However, this high velocity material produces broad features observed at early epochs \citep{Hat99, Maz05}.  It is possible that the detached material could produce absorptions at late-times, but it is difficult to reconcile how this material would recreate the observed relatively narrow and transient features that can last for a few months.

%%%%%%%%%%
% Subsection: Possible answers
%%%%%%%%%%

%%%%%%%%%%
% Subsubsection: Novae Ejecta
%%%%%%%%%%
\subsection{Ejecta from Prior Novae Eruptions}

Due to their variability and transient nature, these late-time absorption features could be the result of the SNe interacting with the surrounding material.  \citet{Pat0706X,Pat11} suggested that \Ias\ may be interacting with dense shells of slow-moving ejecta from prior nova eruptions.  In the single-degenerate \Ia\ model, the white dwarf can experience multiple novae events prior to the final SN explosion \citep{Hac01}.  When this happens, the white dwarf expels material from its surface as clumpy ejecta at velocities of a few thousand km s$^{-1}$.  Such recurrent novae events are thought to increase the white dwarf's mass with each outburst, thus helping it reach M$_{\rm{CH}}$ \citep{Sta85, Liv92, delVal96}.

\citet{Dil12} showed that PTF~11kx exhibited narrow \ion{Fe}{2} 5018, 5169, 5316~\AA\ and \ion{He}{1} absorptions in its day $-$1 to +58 spectra.  They propose that the material likely lies in a ring in the equatorial plane surrounding the SN, like that observed in the 2006 nova of RS Ophiuchi \citep{Hac01,OBr06, Bod06, Bod07}.  The observed delays in the features' appearance could be due to the area surrounding the SN having been evacuated of material by prior novae \citep{Woo06}.

The transient nature of the narrow 4760, 4840, and 5000~\AA\ absorption features we have found might be consistent with a ring of ejected material, where rings of varying thickness will determine the length of time the features are observable in the spectra.  The disparity in epoch that these features are first observed could be the result of systems undergoing a larger number or stronger novae events than others, thus creating larger evacuated regions.  The constant wavelengths of these absorptions could be attributed to the ejecta moving at a constant velocity.

However, there are some problems with this explanation.  \citet{Dil12} suggests that there should also be a ring of CSM which should present as hydrogen emission in the spectra, but none of the \Ias\ in our sample show any evidence of hydrogen emission or absorption.  Also, without knowing the velocity of this ring, determining which species are responsible for the absorptions will be challenging.

%%%%%%%%%%
% Subsubsection: Mass Loss
%%%%%%%%%%
\subsection{Pre-SN Mass Loss Material}

Prior to the SN event in the single degenerate model, the environment surrounding the white dwarf should contain a considerable amount of hydrogen and helium lost from the companion star.  The absorptions at 4840 and 5000~\AA\ could be the result of H$\beta$ and \ion{He}{1} 5016~\AA\ absorption from hydrogen and helium envelopes traveling at $\sim$1300\vel and $\sim$900\vel respectively.  The notch at 4760~\AA\ would not be an absorption in this case, but rather the emergence of weak [\ion{Fe}{3}] emissions around 4750~\AA\ as we previously proposed \citep{Bla16}.

However, a problem in this explanation is the disparity in velocity between the hydrogen and helium shells.  While it is reasonable to think that the hydrogen shell might be at a higher velocity, a difference of roughly 400\vel may be too large to reconcile.  Moreover, the lack of absorption at H$\alpha$ would be expected in addition to H$\beta$, although, a weak red continuum might explain the apparent absence of H$\alpha$ absorption.

%%%%%%%%%%
% Figure: Fe Blob
%%%%%%%%%%
\begin{figure}[!ht]
        \centering
          \subfigure[]{ \label{fig:Blob1}
		\includegraphics[width=0.85\columnwidth]{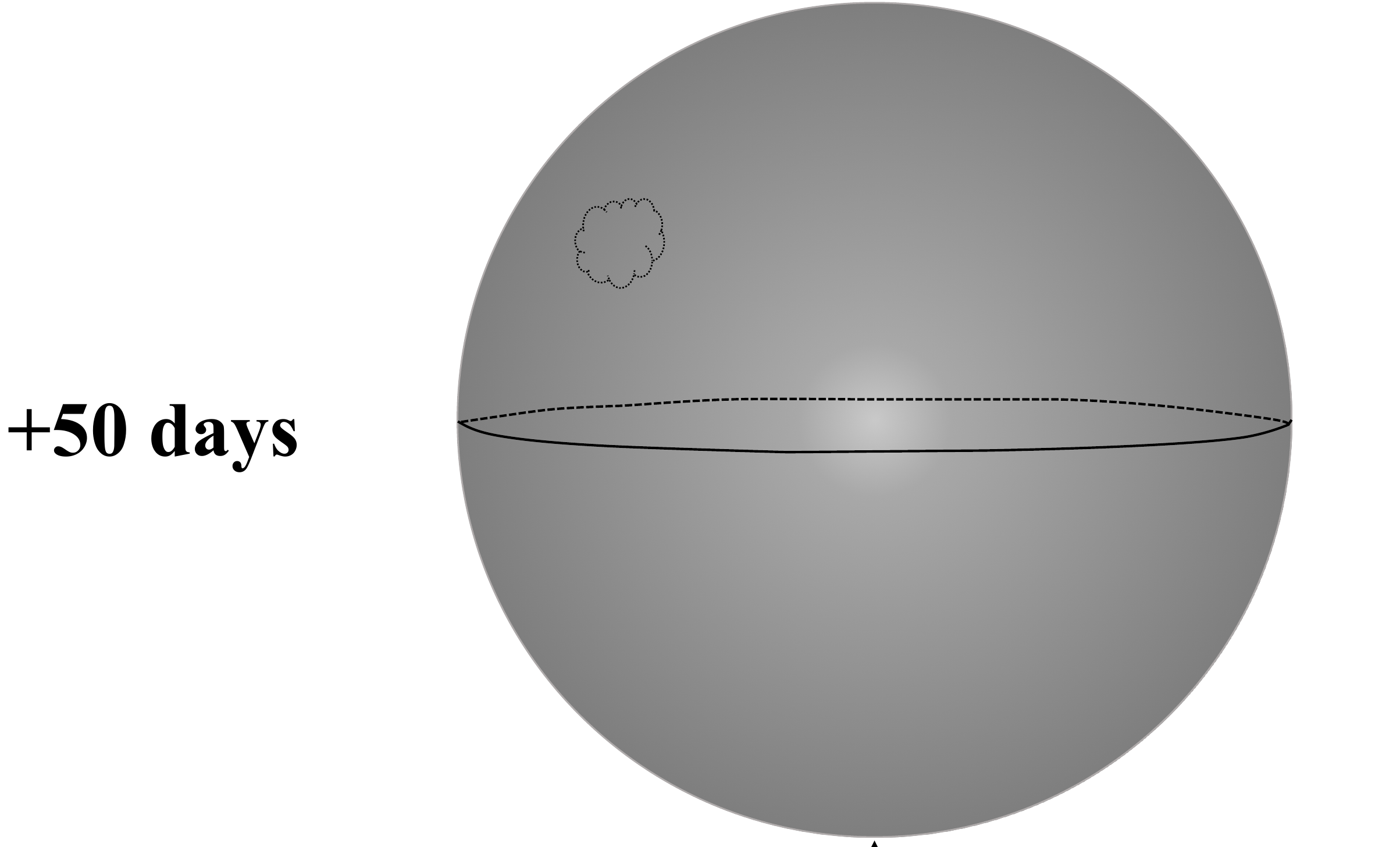}}
          \subfigure[]{ \label{fig:Blob2}
		\includegraphics[width=0.85\columnwidth]{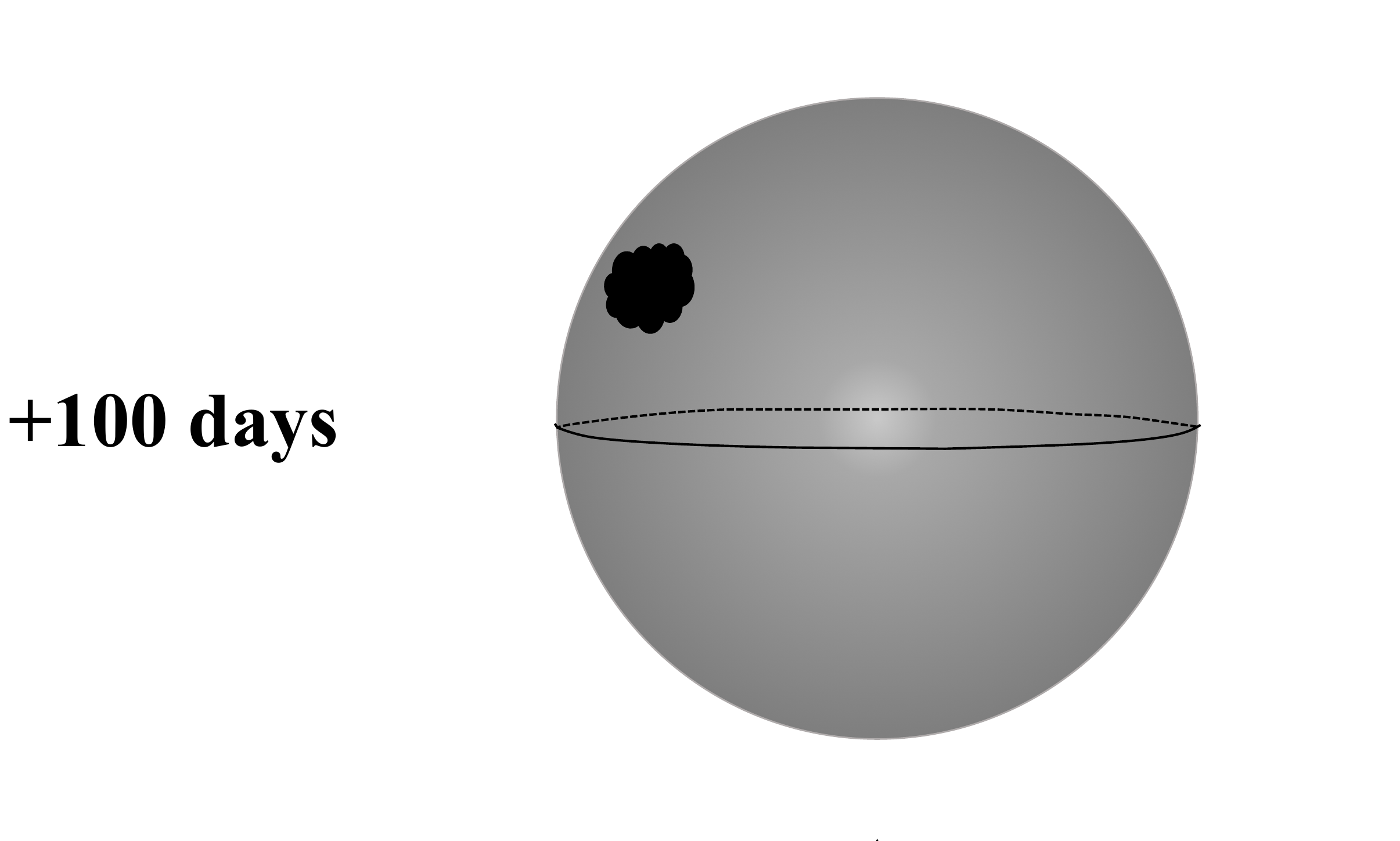}}
          \subfigure[]{ \label{fig:Blob4}
		\includegraphics[width=0.85\columnwidth]{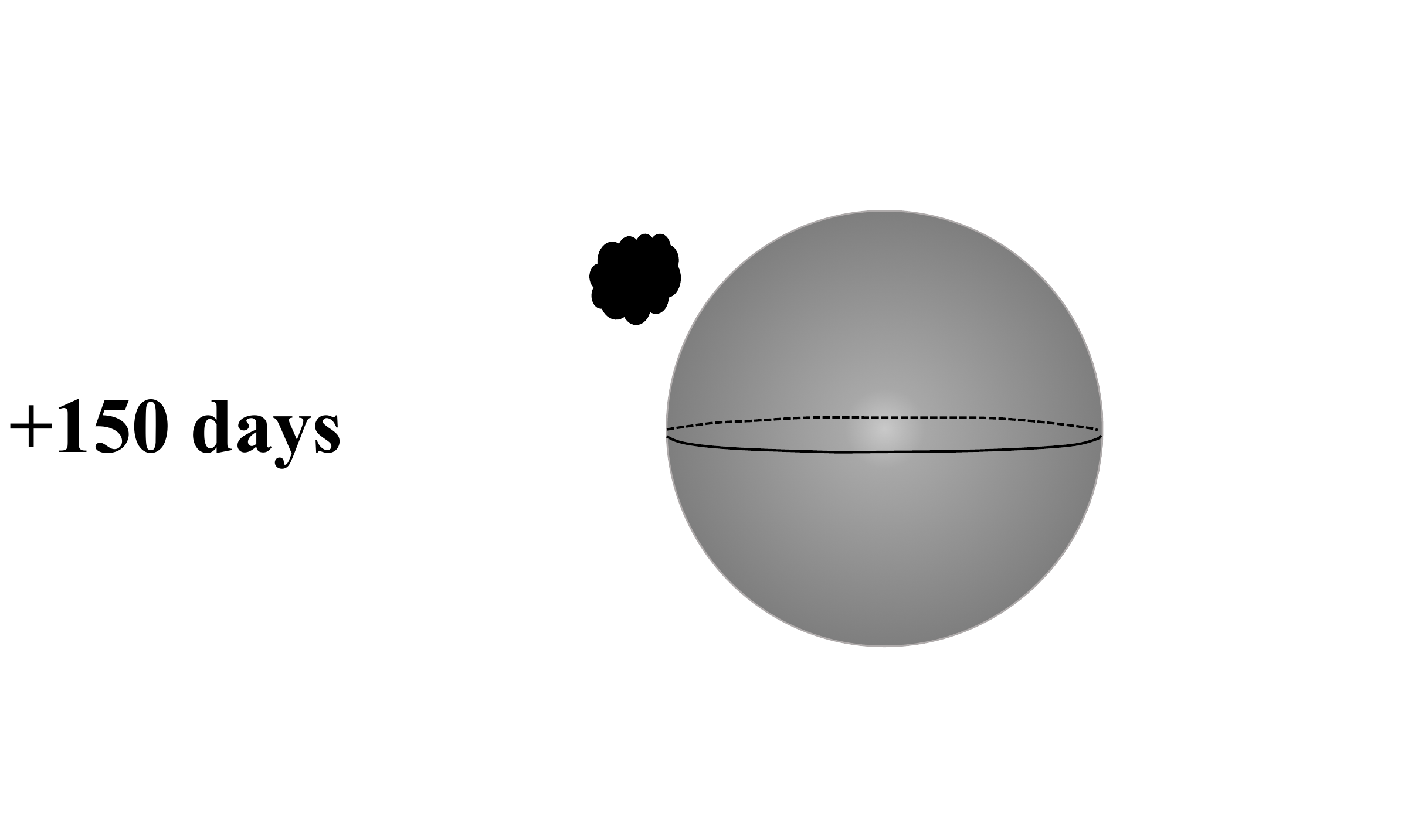}}
        \caption{A cartoon of the scenario where the narrow absorptions are the result of Fe clumps, where the days listed are given as an example. (a) The clump(s) lie below the optically thick, receding photosphere and are not seen in the spectra.  (b) The clump(s) begins to show above the photosphere.  (c) The photosphere has receded enough such that the clump(s) is no longer back-lit and the absorptions disappear from the spectra.}
        \label{fig:Feblob}
\end{figure}
%%%%%%%%%%

\subsection{Clumps of Fe-Rich Ejecta}

Finally, we consider whether these narrow absorption features could be the result of Fe-rich ejecta clumps that become revealed at later times as the photosphere recedes.  Evidence of clumpy Fe-rich ejecta in \Ias\ can be seen in X-ray images of the \Ia\ remnant Tycho (SN~1572) which has spherically distributed but relatively clumpy Si-rich ejecta \citep{Sew83, Van95, Hwa02}.  A knot on the SE border of the remnant has been found to contain mostly Fe-rich material that is chemically distinct from the surrounding ejecta \citep{Yam17}.  \citet{Tse15} suggest that Fe clumps may also be responsible for the protrusions or `ears' seen in Kepler's SNR and SNR G1.9+0.3.

Additional evidence of Fe-rich ejecta clumping comes from the young \Ia\ remnant S Andromeda (SN~1885).  HST images and spectra of this SN remnant indicate Fe-rich ejecta concentrated in four plumes, extending from the remnant center out to 10,000\vel\ \citep{Fes15,Fes17}. 

Such Fe-rich clumps in these young SNRs suggest that Ia explosions can create a few distinct Fe clumps. These might be observable in SNe spectra at late epochs.  Figure \ref{fig:Feblob} is a cartoon of how an Fe clump with velocities of $\sim$8,000~$-$~12,000\vel\ underneath a debris field that creates a pseudo-photosphere of scattered radiation would become increasingly exposed.  An initial large optical depth can explain the delay in the appearance of these features as well as their evolution.  As the pseudo-photosphere becomes increasingly optically thin, the Fe clump emerges and consequently the absorptions strengthen.  Eventually, the effective radius of the photosphere decreases such that the Fe clump no longer absorbs the photospheric flux.

%%%%%%%%%%
% Table: Vel of Fe Clumps
%%%%%%%%%%
\begin{deluxetable}{lrrrc}
  \tablecolumns{5}
  \tablecaption{Blueshifted Velocities of the Fe Clumps \\ For \ion{Fe}{2} 4924, 5018, and 5169~\AA}
  \tablehead{
  \colhead{SN} & \colhead{4760~\AA} & \colhead{4840~\AA} & \colhead{5000~\AA} & \colhead{FWHM (\AA) of} \\
  \colhead{ID}  & \colhead{\vel} & \colhead{\vel} & \colhead{\vel}& \colhead{5000~\AA\ Feature}}
  \startdata
  	1994ae & \nodata & \nodata & 9,000 & 18 \\
	1997do & 11,200 & \nodata & 10,700 & 25 \\
	2000cx & 11,500 & 11,700 & 11,500 & 12\\
	2001V & 10,200 & \nodata &  10,400 & 21\\
	2004bv & \nodata & 9,200 &  8,900 & \nodata \\
	2007sr & 9,400 & \nodata & 9,300 & 23\\
	2011fe & 8000 & 8200 & 8400 & \nodata \\
	2012fr & 9,400 & \nodata &  9,000 & 35\\
	2013dy & 8,300 & \nodata & 8,600 & 12\\
	20014J & 10,300 & \nodata & 10,400 & \nodata\\
	ASASSN-14lp & 10,500 & 10,500 & 10,300 & 13\\
	2017bzc & 9,600 & 9,700 & 9,600 & 12
   \enddata
   \label{tab:Vel}
   \tablenotetext{}{The 5000~\AA\ features in SNe~2004bv and 2014J are too shallow to yield a meaningful FWHM measurement.}
\end{deluxetable}
%%%%%%%%%%

%%%%%%%%%%
% Figure: Vel vs Time
%%%%%%%%%%
\begin{figure}
        \centering
		\includegraphics[width=\columnwidth]{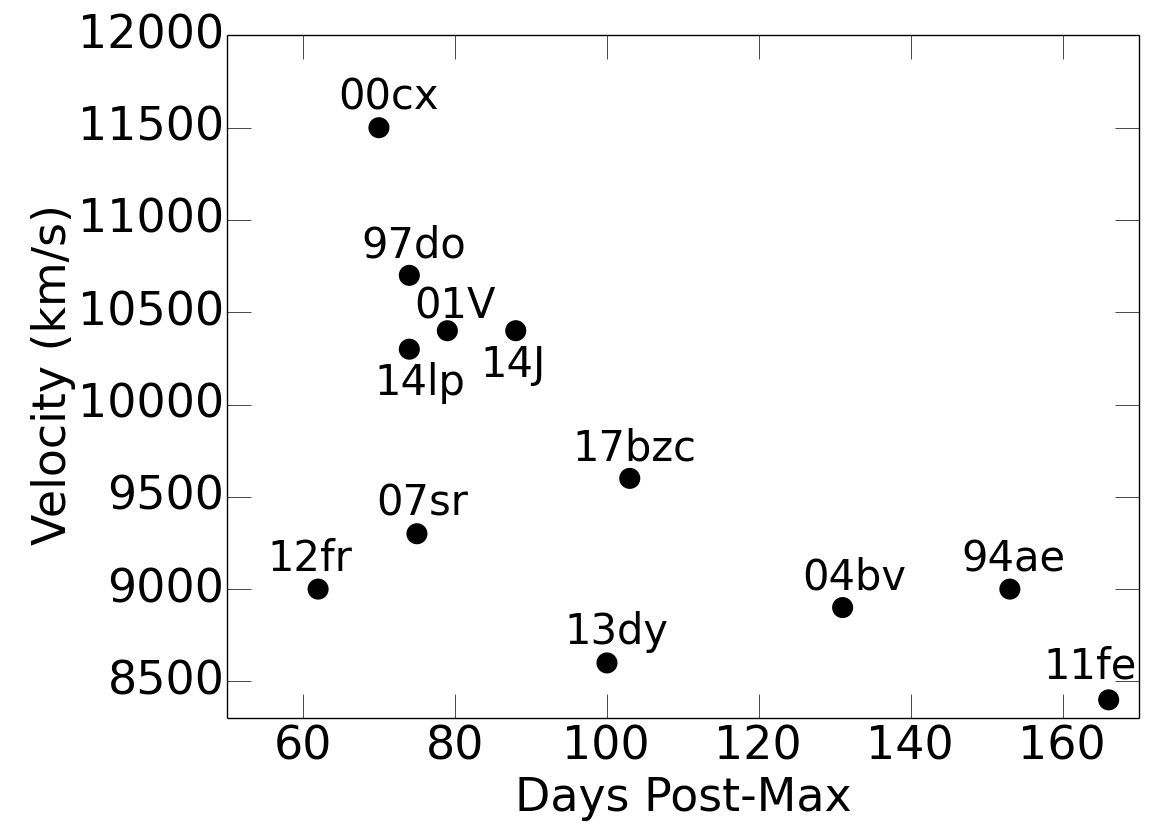}
        \caption{The blueshifted velocity of the 5000~\AA\ feature due to \ion{Fe}{2} 5169~\AA\ in relation to the first appearance of the transient features in the spectra.}
        \label{fig:VelvTime}
\end{figure}
%%%%%%%%%%

It is possible that the transient features discussed here are the result of absorption from permitted \ion{Fe}{2} 4924, 5018, 5169~\AA, which make up multiplet 42 of \ion{Fe}{2}.  If true, then the blueshifted absorption velocities range from roughly 8,000\vel to 11,500\vel across the \Ias\ in our sample.  In our sample SN~2000cx has the largest velocity of 11,500\vel\ and SN~2011fe has the lowest at 8,400\vel.  These velocities are consistent with those reported  for the Fe plumes of SN~1885, which are observed to extend out to 10,000\vel \citep{Fes15}.  Table \ref{tab:Vel} lists the estimated velocities of each feature for the SNe in our sample as well as the measured FWHM of the 5000~\AA\ feature.

In this Fe clump model, observed FWHMs provide lower limits on the velocity range of the clumps due to the small relative absorptions in the 4700 - 5000~\AA\ range.  Using ASASSN-14lp at day +114 as an example, we fit the absorptions with a Gaussian assuming a continuum level roughly equal to the peak of the 4700~\AA\ feature.  This yields FWHM values of 35, 55, and 45~\AA\ for the 4760, 4840, and 5000~\AA\ absorptions, respectively, which translates to an effective clump velocity range of 2000 $-$ 2500\vel.

The observed blueshifted velocities of the 5000~\AA\ absorptions are compared to the approximate epoch that the features first appear in Figure \ref{fig:VelvTime}.  We find a possible trend of higher velocity absorptions found at earlier epochs, supporting the Fe clump model as clumps that are revealed at earlier times should be moving at higher velocities than those that do not appear until later times.  However, we caution that this apparent correlation may simply be due to the lack of high cadence, multi-epoch spectra of \Ias\ after +100 days.

The range of \ion{Fe}{2} clump velocities observed in our sample could be the result of the projected radial velocity of the clump's expansion velocity with respect to our viewing angle.  Assuming that these Fe clumps are moving with a maximum velocity of $\simeq$10,000~$-$~12,000\vel, the observed velocities can be recreated with clumps moving at angles of 10-45 degrees.

In addition, the varying FWHM values of these narrow features seen between objects could be due to a differing clump covering fractions of the SN photosphere 2-5 months post-max. For example, a larger covering fraction might explain how some supernova, like SN~2012fr, show significantly stronger features while a smaller fraction would generate weaker absorptions like that seen in SN~2000cx.  

Of the possible explanations offered to describe the origins of these transient absorption features, we favor the Fe clump model.  Clumps of Fe explain how the absorption features appear simultaneously in the spectra where variations of the projected clump radial velocities result in features with a range in wavelengths across \Ias, while still maintaining the relative spacing between the absorptions.

The transient nature of the absorption features is consistent with a receding photosphere increasingly revealing a discrete clump and the length of time they can be seen is dependent of the clump covering fraction.  Support for this model comes from the observed rise and decline of late-time gamma rays from $^{56}$Co decay around day +60 to +100 in SN 2014J \citep{Die15}, which occurs in nearly the same timeframe and durations as when the narrow 4760, 4840, and 5000  \AA \ features appear in its optical spectra (see Fig.\ 3). This variability of SN 2014J's gamma rays at late-times has been interpreted as possibly due to the exposure of clumps of $^{56}$Co along less occulted line-of-sight as the supernova's cloud of debris expands and evolves (Diehl et al. 2015). This is consistent with an Fe clump model (see Fig.\ 8) and could apply to any SN Ia subtype, thus explaining narrow transient features present in both normal and 91T-like SNe Ia.

If such Fe-clumps exists, their presence and origin could have important implications regarding \Ias\ explosion dynamics. Large Fe-rich clumps possibly associated with Rayleigh-Taylor plumes, like those seen in the remnant of SN~1885 \citep{Fes15}, might inform us regarding the chemical and dynamical properties near the explosion center.  However, X-ray observations of the Fe knot in the Tycho SNR indicate that the knot did not originate from deep within the white dwarf progenitor \citep{Yam17} and that the Fe is likely a nonradioactive form, like \el{54}{Fe} \citep{Wan01}.

%%%%%%%%%%
% Section: Conclusions
%%%%%%%%%%
\section{Conclusions}\label{sec:14lpConc}

A sample of 12 \Ias\ with a total of 58 optical late-time spectra, including the observations of ASASSN-14lp and SN~2017bzc, are found to exhibit transient, narrow absorptions features between 4700-5000~\AA. The presence of such absorptions is unexpected in \Ia\ spectra where only broad, blended features are seen even at late times, well into the nebular phase. The fact that these absorption features are observed in four of twelve bright recent SNe suggests that such absorption features are not an especially uncommon phenomenon.

The specific findings of our study include:

1) Three narrow features can be seen in the spectra of nearly a dozen \Ias\ between +60 and +200 days. Two absorption features are found at $\sim$4840 and 5000~\AA, as well as a distinct notch at $\sim$4760~\AA.  They appear in moderate to high S/N spectra of both normal and 91T-like \Ias, including SN~2011fe.  These are a relatively common phenomenon in \Ias\ as we estimate that roughly 10\% or more of \Ias\ exhibit these narrow absorptions.

2) These absorptions appear in the spectra almost simultaneously after day +60 and disappear some one to three months later. Such transient nature of late-time features has not been observed before and is unexpected in \Ia\ spectra.  Additionally, the features do not exhibit a shift in wavelength, unlike the adjacent blue spectral features, and they maintain roughly similar wavelengths, ranging up to $\pm$20~\AA, across the sample of \Ias.  The relative spacing between these absorptions is the same across the SNe in our sample to within measurement error.

3) Because these three absorptions evolve together as the features strengthen and then eventually disappear from late-time \Ias\ spectra, they likely have a common source.  While they may not all be from the same element, they likely originate from the same environment.

4) These features could be absorptions due to previous mass loss events or shells of ejecta from repeated novae events prior to the SN explosion.  However, we favor an alternative possibility where these features are the result of Fe clumps revealed by a receding photosphere, where the observed absorptions seem to match with the \ion{Fe}{2} 4924, 5018, 5169~\AA\ features.

\smallskip
The puzzling appearance and disappearance of narrow absorptions several months after maximum light, affecting even the profile of the prominent, late-time 4700~\AA\ feature, may have important implications on our understanding of the environment surrounding a \Ia, or indicate the late-time emergence of discrete clumps of Fe-rich ejecta. In either case, the presence of these unexpected features in a variety of \Ias\ subtypes, including the prototypical SN~2011fe, deserve to be understood.

%%%%%%%%%%
% Section: Acknowledgements
%%%%%%%%%%
\acknowledgments
The authors wish to thank the referee for providing helpful comments, K. Zhang for providing spectra of SN~2014J, and the staffs of MDM and SALT observatories for their assistance in making these observations possible.
This research was supported in part by a Fellowship from Dartmouth's School of Graduate and Advanced Studies and RAF's Archangel Research Program.  This work was made possible by contributions to the Open Supernova Catalog (OSC; \citealt{OSC}) and the Weizmann Interactive Supernova Data Repository (WISeREP; \citealt{Yar12}).

\bibliographystyle{apj}

\end{document}